%% file: Kings_arxiv.tex
\documentclass{article}
\usepackage{epsfig}
\usepackage{colortbl}
\usepackage{ulem}
\usepackage[left]{lineno} 
\usepackage{amsmath, amssymb, amsfonts}

\def\@xthm#1#2{\@beginassumption{#2}{\csname the#1\endcsname}{}\ignorespaces}
\def\@ythm#1#2[#3]{\@opargbeginassumption{#2}{\csname the#1\endcsname}{#3}\ignorespaces}%
\def\@beginassumption#1#2#3{\par\addvspace{8pt plus3pt minus2pt}%
              \noindent{\csname#1headfont\endcsname#1\ \ignorespaces#3 #2.}%
              \csname#1font\endcsname\hskip.5em\ignorespaces}
\def\@endassumption{\par\addvspace{8pt plus3pt minus2pt}\@endparenv}
\usepackage{tikz}
\usetikzlibrary{shapes,arrows}

\usepackage{multirow}

\usepackage{hyperref}

\usetikzlibrary{shapes,arrows,decorations.markings}
\usetikzlibrary{decorations.pathmorphing}

\tikzstyle{block_long} = [rectangle, draw, fill=blue!20,
    text width=10.0em, text centered, rounded corners, minimum height=3em]
\tikzstyle{block_medium} = [rectangle, draw, fill=blue!20,
    text width=6.0em, text centered, rounded corners, minimum height=3em]
\tikzstyle{block} = [rectangle, draw, fill=blue!20,
    text width=3.0em, text centered, rounded corners, minimum height=3em]
\tikzstyle{line} = [thick, draw, dashed,  -stealth']

\usepackage{mathrsfs,bbm}
\usepackage[american]{babel}

\newcommand{\cA}{\mathcal{A}}

\newcommand{\cD}{\mathcal{D}}

\newcommand{\cP}{\mathcal{P}}

\newcommand{\RR}{\mathbb{R}}

\newcommand{\bomega}{\boldsymbol{\omega}}

\newcommand{\bx}{{\boldsymbol{x}}}

\newcommand{\bu}{{\boldsymbol{u}}}
\newcommand{\bv}{{\boldsymbol{v}}}
\newcommand{\bw}{{\boldsymbol{w}}}
\newcommand{\bk}{{\boldsymbol{k}}}

\newcommand{\bq}{\boldsymbol{q}}

\newcommand{\bU}{{\boldsymbol{U}}}
\newcommand{\bQ}{{\boldsymbol{Q}}}
\newcommand{\bbf}{\boldsymbol{f}}

\def\va{\raise 2pt\hbox{,}}

\def\cG{{\cal G}}

\def\cL{{\cal L}}

\def\cA{{\cal A}}

\def\cD{{\cal D}}

\newcommand{\p}{\partial}

\input epsf
\begin{document}


\markboth{M.~Dolfin, L.~Leonida}{From classical to active particles methods}

%
%

\title{From classical to active particles:\\ mathematical tools\\ for social dynamics and behavioural economics.}

\author{Marina Dolfin$^{(1)}$ and Leone Leonida$^{(2)}$} 

\date{$^{(1)}$King's College London, London, U.K., and University of Messina, Italy.\\
marina.dolfin@kcl.ac.uk\\
\vskip.1cm
$^{(2)}$King's College London, London, U.K.\\
leone.leonida@kcl.ac.uk\\
}

\vskip1truecm

\maketitle

\begin{abstract}
This essay provides a critical overview of the mathematical kinetic theory of active particles, which is used to model and study collective systems consisting of interacting living entities, such as those involved in behavior and evolution. The main objective is to study the interactions of large systems of living entities mathematically. More specifically, the study relates to the complex features of living systems and the mathematical tools inspired by statistical physics. The focus is on the mathematical description of these interactions and their role in deriving differential systems that describe the aforementioned dynamics. The paper demonstrates that studying these interactions naturally yields new mathematical insights into systems in the natural sciences and behavioral economics.
\end{abstract}

\noindent\textbf{Keywords:} Life, active particles, interactions, complexity, collective dynamics.\\
\textbf{AMS Subject Classification:} 82D99, 91D10


\section{Motivations and Plan of the Paper}\label{Sec:1}

This paper offers a critical analysis of the study of interactions in large systems of living, and therefore behavioral, individual entities.
First, we describe the physical properties of the systems under consideration. Next, we justify studying interactions within the mathematical framework of differential systems of active particles. Finally, we offer a critical analysis focusing on the intriguing yet challenging perspective of studying the collective dynamics of large systems of interacting living entities.

In particular, we focus on differential systems developed within frameworks that account for the distinctive features of living systems. According to~\cite{[BBD21]}, these key features include the ability to express individual and heterogeneously distributed  strategies; engage in both proliferative and destructive interactions; and evolve through mutation, potentially followed by selection. In what follows, we discuss the central role of interactions in modelling the collective dynamics of living systems.

Such dynamics exhibit common features shared by any specific class of systems, as well as particular features that account for the specificity of each system. Examples include multicellular systems, animal swarms, and human crowds, as well as pseudo-living systems, such as programmed robots or drones. This study also considers the dynamics of individuals within social, political, economic, and other interest groups in our society.

Collective dynamics can be described by differential systems suggested by mathematical approaches, which are sometimes associated with the ambitious definition of a ``mathematical theory''. We focus particularly  on the kinetic theory of active particles, as reviewed in~\cite{[BBD21]} and~\cite{[BBL25]}. We also explore  parallel methods, such as Fokker-Planck-Boltzmann theory (see~\cite{[PT13]}).

These approaches are characterized by their ability to derive a differential system that translates interactions on the microscopic scale into a description of the system's collective motion. The mathematical description of these interactions is a key aspect and the specific topic addressed in our paper. We first review and critically assess recent studies in this area, then examine how these models can be applied to various case studies.

An important conceptual tool is game theory. In particular, we highlight Nash's foundational contributions~\cite{[NASH51]} and~\cite{[NASH96]} with applications to economics. The book by Nowak and May~\cite{[NM01]} is a milestone for the application of game theory in life sciences.  For the theory of mean field games we refer to~\cite{[LL06A],[LL06B]}, and~\cite{[LL07]}, for emphasis on applications, see~\cite{[BCL21],[BBLL24]}, and~\cite{[Bar25]}.

\begin{figure}[t!]\label{steps}
\begin{center}\scalebox{0.80}{
\begin{tikzpicture}[node distance = 2.8cm, auto]
\node [block_long]  (complex) {1. Complexity\\ features, \\ behavioral variables \\ and self-organization};
 \node [block_long, below of=complex] (structure) {2. Selection of \\ differential structures:\\ in the framework of \\ kinetic theories };
  \node [block_long, below of=structure] (model) {5. From interactions \\ to the derivation of \\ differential models};
 \node [block_long, left of=structure,xshift=-3.2cm] (micro) {3. Modelling \\individual based \\interactions};
 \node [block_long, right of=structure,xshift=3.2cm] (social) {4. Modelling\\ external  actions\\ on individual entities};
\node [block_long, below of=model] (development) {6. Applications:\\ analytical and\\ computational \\challenges};
 \draw [->] [line width=.5mm, blue, ->] (complex.west) -- (micro.north);
 \draw [->] [line width=.5mm, blue, ->] (complex.east) -- (social.north);
 \draw [->] [line width=.5mm, ->] (complex.south) -- (structure.north);
 \draw [->] [line width=.5mm, ->] (social.west) -- (structure.east);
 \draw [->] [line width=.5mm, blue, ->] (social.south) -- (model.east);
 \draw [->] [line width=.5mm, ->] (micro.east) -- (structure.west);
 \draw [->] [line width=.5mm, blue, ->] (micro.south) -- (model.west);
 \draw [->] [line width=.5mm, ->] (structure.south) -- (model.north);
 \draw [->] [line width=.5mm, ->] (model.south) -- (development.north);
\end{tikzpicture}}
\end{center}
\begin{center}
\caption{Sequential steps of the modelling approach}
\end{center}
\end{figure}
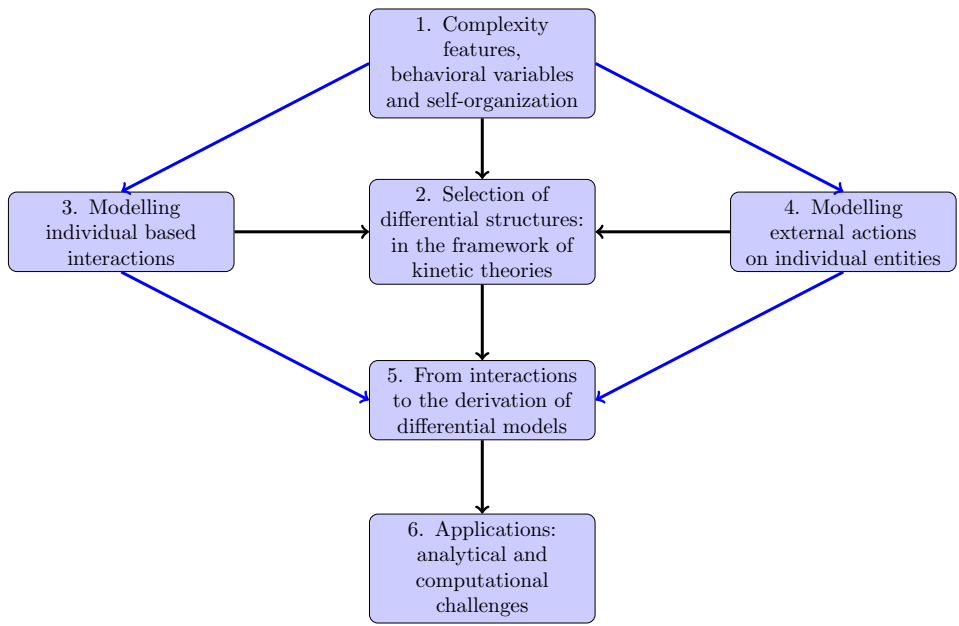

The flow chart in Fig.\ref{steps} illustrates the sequential steps of the modelling approach, as proposed in \cite{[BD06]}, and highlights how our paper, focused on modelling interactions, contributes to this framework. The exposition begins with conceptual foundations and progressively introduces technical details, offering guidance for readers interested in the analytical aspects. In most cases open access articles are selected so that the reader can rapidly recover mathematical tools.

The discussion then transitions to applications.  In fact, the study can contribute to real-world applications, and, in particular, to mathematical development of scientific machine learning  methods towards the study of collective dynamics of large systems of interacting living entities, see~\cite{[BDL24]}.

The systems under consideration are living, hence \textit{behavioral} and \textit{complex}. We adopt the interpretation by Herbert. A.~Simon (Nobel awarded), see~\cite{[Simon1965],[Simon2019]} and~\cite{[DCK24]}. We specifically refer to the following sentence:
\begin{quote}
\textit{Roughly by a complex system I mean one made up of a large number of parts that interact in a non-simple way. In such systems, the whole is more than the sum of the parts, not in an ultimate metaphysical sense, but in the important pragmatic sense that, given the properties of the parts and the
laws of their interaction, it is not a trivial matter to infer the properties of the whole.}
\end{quote}

The motivations for our study have been outlined. We now present the structure of the paper, which is planned as follows:

\vskip.2cm \noindent  Section 2 offers a brief overview of the mathematical theories considered in this paper. It introduces key preliminary concepts that form the foundation for the mathematical modelling of interactions, which is the focus of the subsequent sections. In particular, we introduce the celebrated Boltzmann equation and the kinetic theory of active particles. Related methods, that have been developed in parallel, are also briefly discussed for comparison.

\vskip.2cm \noindent
Section 3 presents a critical analysis of the complexity inherent in collective living systems. This analysis serves as a foundation for studying interactions, as it introduces the conceptual frameworks necessary to describe individual-based dynamics is the central focus of our paper. Understanding this complexity is also essential for selecting an appropriate mathematical framework to guide the modelling approach and, ultimately, the study of interactions.

\vskip.2cm \noindent Section 4 focuses on modelling interactions within the framework of kinetic theory methods. The conceptual differences between interactions in living systems and those among classical particles are critically examined. This study is referred to the context of modelling real-world systems. Both models with spatial dynamics and those in the space-homogeneous setting are considered. This section also reports about some applications, demonstrating how interaction modelling at the microscopic scale can be applied to specific case studies.

\vskip.2cm \noindent Section 5 offers a forward-looking perspective on the research approaches that emerged from the critical analysis presented in previous sections. It focuses on systems in the social sciences and economics related to collective learning and decision-making.  The main objective is to identify systems that can describe how the brain and socio-economics contribute to the dynamics of these complex systems.

\section{Theories of the kinetic theory of classical and active particles}\label{Sec:2}

This section provides an initial description of the class of systems on which our research is focused. Specifically, we examine the collective dynamics of classical or active particles. We employ mathematical theories that transfer dynamics from the individual scale to the collective scale via differential systems. This approach allows us to study interactions within general mathematical and physical frameworks while considering the key features of each specific system.

First, we briefly consider classical particles within the framework of the kinetic theory of rarefied gases and the Boltzmann equation. Foundational references include Boltzmann's original work~\cite{[Boltzmann]}, as well as more recent treatments such as~\cite{[CIP93]} and~\cite{[Kogan]}.  Next, we focus on systems of living matter, especially large systems of interacting living entities. For this class of systems, a key reference is the open-access review~\cite{[BBD21]}, which offers a critical overview of the Kinetic Theory of Active Particles (KTAP).  These topics are presented in the next two subsections.  The final subsection  some recent summarizes developments in KTAP theory.

The focus of our paper is the dynamics of living systems, particularly active particle methods.  Although our focus is on living matter, references to classical theories offer valuable insights into its complexity. We also briefly review related approaches, such as the Fokker-Planck-Boltzmann theory, which has been applied to model the collective dynamics of active particles and behavioural dynamical systems (see, for example,~\cite{[PT13]}). These systems are somewhat related to the theory of behavioural swarms, see reference~\cite{[FLO25]}.

\subsection{Differential structures for the dynamics of classical particles}\label{Sec:2.1}

Consider a large system of  \textit{classical particles}, modeled within the framework of the kinetic theory of rarefied gases or, more broadly, statistical physics , see~\cite{[CIP93],[Kogan]}. In classical theory, the dynamics are described by the \textit{Boltzmann equation}, which was originally proposed for a single population. In the case of classical particles, the microscopic state is then given only by position and velocity, while the distribution function is $f = f(t, \bx, \bv)$. While there are some analogies with systems of living entities, often referred to as active particles, there are also fundamental differences that distinguish the latter.

The derivation of the equation is obtained by balancing the particles in the elementary volume of phase space
$[\bx, \bx + d\bx] \times [\bv + d\bv]$, and can be formally obtained by equating the flux due to transport with the net flux due to interactions. In the absence of external forcing, the mathematical structure is as follows:
\begin{equation}\label{transport}
\big(\p_t + \bv \cdot \nabla_\bx \big) f(t, \bx, \bv) = G(f,f)(t, \bx, \bv) - L(f,f)(t, \bx, \bv),
\end{equation}
where the left term defines the transport and the right term defines the net flow. $G$ and $L$ denote the so-called \textit{gain} and \textit{loss} terms and correspond to the inlet particles and the outlet from the said volume, respectively. The interactions are non-dissipative, i.e. mass and momentum are conserved.

The derivation of kinetic theory models of classical particles (c-particles)  needs some heuristic assumptions:
\begin{enumerate}
\vskip.2cm \item The probability of occurrence of interactions involving more than two particles is negligible compared to the probability corresponding to binary encounters.

\vskip.2cm \item The asymptotic pre-interaction velocities of two molecules are not correlated as well as their post-interaction velocities. This hypothesis is known as the \textit{molecular chaos assumption} and implies that the joint particle distribution functions of the two interacting particles  can be factorized, i.e.  given by the product of the distribution functions of the interacting particles.

\vskip.2cm \item Although particles interact at a distance, however small, it is assumed that the interacting distribution functions is that in the localization of the distribution of the \textit{test particle} which is representative of the whole system.
\end{enumerate}

\subsection{Differential structures for the dynamics of active particles}\label{Sec:2.2}

Let us now focus  on the kinetic theory of \textit{active particles}, a-particles for short. The theory naturally leads to different interacting systems, called \textit{functional subsystems} (FSs). We refer to ~\cite{[BBD21]} and to the reviews~\cite{[BBD21]} and~\cite{[BBL25]}. The state of the whole system is defined by the one-particle distribution function for a system of $m$ FSs:
\begin{equation}\label{2.1}
f_i = f_i(t, \bx, \bv, \bu), \hskip1cm i = 1, \ldots, m,
\end{equation}
where the microscopic state of the a-particles is given by their position $\bx \in \RR^3$, velocity $\bv \in \RR^3$, and activity $\bu \in D_{\bu}$, while functional subsystems are denoted by the subscript $i$. In general $\bu$ is a vector, but in most applications it is a scalar variable.
A reference to statistical physics is important, and we must emphasize the conceptual differences in order to avoid naively transferring properties typical of the theory of c-particles to the dynamics of a-particles.
The derivation of the differential structure for systems of a-particles presents some analogies. In fact, it is obtained by a balance of particles in the elementary volume of the space of microscopic states $[\bx, \bx + d\bx] \times [\bv + d\bv] \times [\bu + d\bu]$. The theory also proposes mathematical structures that include proliferative and/or destructive interactions, as well as, interactions that allow migrations across functional systems. The following  formal structure refer to the collective dynamics:
\begin{eqnarray}\label{MultiFS}
&& \partial_t f_i(t, \bx, \bv, \bu) + \bv \cdot \nabla_\bx \, f_i(t, \bx, \bv, \bu) = J_i[\bbf]=  \nonumber\\[3mm]
&& \hskip1cm  \bigg((\cG_i[\bbf] - \cL_i[\bbf])(t, \bx, \bv, \bu) + (\cP_i[\bbf] - \cD_i[\bbf])(t, \bx, \bv, \bu) \bigg),
\end{eqnarray}
where the terms $\cG_i = \cG_i[\bbf]$ and $\cL_i = \cL_i[\bbf]$ represent the gain and loss, respectively, of the number of active particles  entering and leaving the elementary volume $d\bx d\bv d\bu$ of the microscopic state space at time $t$. Similarly $\cP_i[\bbf]$ and $\cD_i[\bbf]$ represent the gain and loss of a-particles due to proliferative and destructive interactions.

We refer to~\cite{[BBD21],[BBL25]} for the detailed expression of the operators on the right-hand side of Eq.~(\ref{MultiFS}).

Before proceeding to the modelling of interactions in Section 3, we outline some key conceptual differences between the two underlying differential structures. The following points also clarify the rationale behind the notations adopted throughout the paper.
\begin{itemize}
\vskip.2cm \item \textit{Bounds on the microscopic variables:} The Boltzmann equation deals with particles treated as point masses, while the  the velocity modulus, which we will call \textit{speed},  is allowed to reach an infinite value. In the case of living matter, the particles have a finite dimension and the velocity modulus has a limit, which can be denoted by $v_M$. Similarly, the activity has a minimum and maximum physical value, in the scalar case $u_m$ and $u_M$, respectively.  These peculiarities suggest using   dimensionless variables for the  speed and the velocity.

\vskip.2cm \item \textit{Nonlinearity of interactions:} The interactions in KTAP theory are nonlinear and nonlocal. This means that the outcome of the interactions depends not only on the microscopic state of the interacting active particles, but also on their distribution functions, i.e. their statistical state. Square brackets, i.e. $[\bbf]$, have been used to indicate that the output of interactions involves the dependent variables, i.e., the distribution functions that describe the overall state of all functional subsystems.

\vskip.2cm \item  \textit{Multi physical systems:} An additional technical difficulty is that the theory is applied to multi-physics problems. This means that the so-called functional subsystems are characterized by different physical properties. This feature requires special attention when modelling the interactions. In some cases, a vector activity variable is necessary to model different components of the social behaviors of the system.
\end{itemize}

\subsection{Critical analysis and selection of a differential structure}\label{Sec:2.3}

In the following sections, we return to KTAP theory, as it offers a coherent framework for capturing the complex features of living systems, as outlined at the beginning of Section 1. Our focus will then shift to the modelling of interaction dynamics. For the sake of completeness, we also briefly mention alternative methods that merit attention and provide relevant bibliographic references.

One topic to be considered is the Fokker-Plank-Boltzmann theory, by which Boltzmann-type kinetic models are first constructed for interacting multi-agent systems, and then a Fokker-Plank-type equation is derived by quasi-invariant asymptotic procedures to capture the long-term behavior of the systems. This theory was motivated by the applications, see\cite{[CPT05],[Toscani06]}, and formalized in the book~\cite{[PT13]}, where the interested reader can also find an excellent presentation of computational methods for kinetic equations.

The Fokker-Plank-Boltzmann method follows the kinetic description of interactions in the spirit of Boltzmann's ideas. More precisely, it is based on prescribed microscopic laws of {\it binary} interactions, depending on the specific case study object of the modelling approach. This theory has a rich record of applications mainly to the modelling and study of complex social systems; for examples of applications see~\cite{[FPTT12],[FPTT17],[FPTT20],[ZBT21]}. Additional applications are reviewed in~\cite{[PT13]}. An excellent review on computational tools in kinetic theory is given in reference~\cite{[DP14]}. Other parallel methods have been inspired to methods of statistical physics, see reference~\cite{[HEL10]}.

Collective dynamics can be described by means of differential systems proposed by the so-called \textit{mathematical theory of behavioral swarms}, which leads to the derivation of a differential system that transfers the description of the interaction at the microscopic scale to the description of the collective motion of the system. This approach shows important differences with respect to kinetic theory. In fact, the a-particles are deterministically individually identified. The microscopic state is different for each a-particle and includes the activity variable in addition to position and velocity.

For the origin of the behaviorally heterogeneous approach, see~\cite{[BHLY24]}, and for further development of the concept of swarm intelligence and application to crowd dynamics, see~\cite{[BHLY24]}, see also~\cite{[GKLY24]}. The study is inspired by the dynamics of c-particles in the mathematical theory of swarms, in particular the seminal work of Cucker and Smale, see~\cite{[CS07]}, which has been technically modified for applications in the modelling of financial markets, see~\cite{[BCLY17],[BCK19]}.
All of the above theories have some connection to the deterministic agent methods, see\cite{[FR21],[GALAM],[Terna]}.

\section{Complexity of the collective dynamics of living systems}\label{Sec:3}

We are interested in describing how the a-particles interact within the mathematical theory of active particles. As a preliminary consideration, it is important to note that  the activity may be difficult to express in terms of physical quantities. This challenge motivates the  use of dimensionless variables, such as the ratio between $v$ and $v_M$, while the geometric space variables are related to a characteristic dimension $\ell$ of the system, such as the diameter of the domain in which the particles move. The reasoning for activity is different, as it can be pragmatically related to the maximum observed value $u_M$, resulting in a dimensionless activity variable defined on the interval  $[0,1]$.

Technically, we can use the following:
$$
\hbox{For the activity} \hskip.5cm \frac{u - u_m}{u_M - u_m}, \hskip.5cm \hbox{and for the velociy}  \hskip.5cm   v\bomega=\frac{v_r}{v_M} \, \bomega,
$$
where $v_r$ is the real velocity, $v_M$ is the maximal possible velocity in real flow conditions and $\bomega$ is the unit vector that identifies the direction of the velocity.

We do start by mentioning  the conceptual difficulty for a mathematics of living systems which has been studied by various authors. In particular, we mention the references~\cite{[Herrero],[MAY],[Nowak],[Reed]}. These essays put in evidence the current absence of a background physics of living system and the need of a search for such theory as a first step toward a mathematics for living systems.

A key theme is the complexity of collective and living systems. This is one of the key topics considered  by the Nobel lecture by Giorgio Parisi~\cite{[Parisi23]} which has guided our study of this topic, whose relevance is enlightened  the following books:

\vskip.2cm \noindent --  Erwin Schroedinger proposed the search for a physics of living systems in his seminal book, which contributes to identifying the multiscale properties of biological matter (see reference~\cite{[ESC1944]}).

\vskip.2cm \noindent --  Ilia Prigogine and  and Isabelle Stengers, a philosopher, proposed a novel and ambitious toward the interpretation of ability of living systems to self-organize, see reference~\cite{[PS1984]}.

\vskip.2cm \noindent -- David C.~Krakauer~\cite{[DCK24]}  provided an excellent framework for this philosophical topic, see also reference~\cite{[KBOFA20]}. In particular, it links concepts to history. The search for a unified theory of life is proposed by C.~Cleland,  see reference~\cite{[CLE19]}.

\vskip.2cm \noindent -- Ernst Mayr has proposed a philosophical theory of the dynamics of evolution (see reference~\cite{[MAYR]}). According to this theory, the dynamics of living systems evolve over time, and evolution is driven by selection. In the search for mathematical tools to describe immune competition at the cellular level, mathematics encountered the theory of evolution. 

\vskip.2cm After conducting a literature review, we must mention Herbert A. Simon's philosophical theory of the artificial world (see references~\cite{[Simon1965]} and ~\cite{[Simon2019]}) and its mathematical interpretation (see \cite{[BE24]}). According to Simon's theory, interaction rules evolve based on environmental dynamics and the actions of interacting entities (see also \cite{[Sim55]}).

Motivations for applications and for a complexity analysis as a preliminary step to modelling have been proposed in~\cite{[Ball12]}.
Complexity studies and features have been considered in various applications to real-world systems, such as vehicular traffic, see~\cite{[PH71]} and~\cite{[PF75]}, as well as more recent developments~\cite{[CDF07],[HEL01],[KW1999]}. Other developments  consider the drivers' ability, see~\cite{[BDF17]} and multi-lane dynamics, see reference~\cite{[zagour]}. For crowd dynamics we refer to~\cite{[BGQR22]}, see also~\cite{[BGO19]}. The approach has been further developed to describe space dynamics linked to  biological dynamics of cells, see references~\cite{[BDZ25],[CKS21],[CDS23]} and~\cite{[TM25]}.  Other applications have been focused on the modelling the immune competition in epidemics, see~\cite{[B3EPT]} (see, for the biology of the immune system, reference~\cite{[MF19]}).

Having outlined the physical differences between the dynamics of classical and active particles, we now turn our attention to the study of interactions. We highlight several considerations that, based on current scientific understanding, should be taken into account in the development of new mathematical tools for modelling interactions and collective dynamics.

\begin{itemize}
\vskip.2cm \item The physical interpretation and derivation of differential systems intended to describe collective dynamics can be applied to various types of living systems, including genes, cells, humans, and even aggregated entities such as corporations, political parties, and various others. Both the physical and philosophical interpretations should follow common principles, aiming to establish a general mathematical framework applicable across a wide range of living systems.

\vskip.2cm \item It is difficult, arguably impossible, to find a general theory of causality principles for living matter. Therefore, according to~\cite{[BBD21]} for the KTAP approach, we follow the idea that the first step is to derive a general mathematical structure suitable for capturing the main features of living systems.  Specific mathematical models are then developed by incorporating the mathematical description of interactions into this structure. These can be obtained by a phenomenological interpretation of each case study under consideration, somewhat consistent with the philosophical theory of life, see~\cite{[CLE19]}.

\vskip.2cm \item Lee Hartwell, Nobel laureate, emphasizes that living systems keep their state far from equilibrium according to their struggle to stay alive, see also~\cite{[HART99],[HART01]}. Indeed, this is different from classical particles, which show a tendency to equilibrium configurations, such as the Maxwellian distribution, which depends on the local macroscopic quantities acting as parameters, see~\cite{[Kogan]}, and~\cite{[CIP93]}.

\vskip.2cm \item The dynamics of the interactions takes place in two steps. The first step is the learning of the overall state of the system of active particles, while the second step for the KTAP approach is a behavioural decision making, by which each active particle modifies the state.  A mathematical theory of the learning dynamics has been developed in~\cite{[BDG16],[BDG16B]}, see also~\cite{[BD19],[CG15]}.

\vskip.2cm \item The a-particles are described in terms of probabilities. As shown in the literature, their modelling can leverage theoretical tools from game theory within a stochastic framework, where interacting entities are represented by probability distributions.  A technical difficulty is that while the approach can be applied to small numbers of particles, non-trivial computational problems arise when dealing with large numbers of particles.

\vskip.2cm \item It is essential to account for multiple interactions, along with their inherent nonlinearity, non-locality, and irreversibility. Additionally, it is important to understand how nonlinearity can generate non-symmetric features in the interactions.

\vskip.2cm \item  An important topic to consider is the concept of \textit{sensitivity domain} in the so-called \textit{topological interactions} conjectured by Giorgio Parisi's team, see~\cite{[BCC08]}, see also~\cite{[Cavagna2008],[Cavagna2008B]} and~\cite{[BCW12]}.

\end{itemize}

\section{Interactions of active particles}\label{Sec:4}

This section focuses on the first of the two main topics of this paper, i.e. a review and critical analysis of the modelling of interactions for the approach of the kinetic theory of active particles. This preliminary study leads to the identification of the physical and conceptual differences between the dynamics of classical and active particles and, subsequently, to the study of the interactions of active particles with emphasis to nonlinearity and asymmetries.

The above topics are covered in the following subsections that deal with:

\vskip.1cm \noindent -- Basic concepts on  the interactions of classical particles.

\vskip.1cm \noindent -- The interaction dynamics of active particles and  symmetry analysis.

\vskip.1cm \noindent -- Critical analysis.

\subsection{Interactions of classical particles}\label{Sec:4.1}

Consider first the \textit{classical kinetic theory}, specifically the Boltzmann equation; see~\cite{[CIP93],[Kogan]}. We also refer to some concepts reported in the original literature~\cite{[Boltzmann]}. Interactions, often called collisions, are binary and, for elastic encounters,  preserve mass, momentum and kinetic energy. The assumption of binary interactions corresponds to the regime of \textit{rarefied gases}, in which the probability of multiple interactions is supposed to be negligible.

If we consider the interaction between two particles with velocities $\bv$ and $\bv_*$ and denote  the post-interaction velocities by a prime, i.e. $\bv'$ and $\bv_*'$, the above mentioned conservation equations correspond to the following equations:
\begin{equation}\label{CCPs}
\begin{cases}
\displaystyle  \bv + \bv_* = \bv' + {\bv}_*   \\[3mm]
\displaystyle  v^2 + v_*^2 = v'^2 + {v'_*}^2,
 \end{cases}
\end{equation}
where the velocity modulus $v$ will be called \textit{speed}, while the interaction rate depends on the \textit{relative velocity} $\bq = \bv - \bv_*$.

System Eq.~(\ref{CCPs}) corresponds to four scalar equations, which are not sufficient to compute the six scalar components of the post-interaction velocities. The remaining two degrees of freedom are fixed by assuming elastic, spherically symmetric interactions, which can be parameterized by a unit vector along the line of centers at impact.

The result is as follows:
\begin{equation} \label{collision-bis}
\displaystyle{\bv' = \bv -\bk (\bk \cdot \bq)},  \hskip.5cm \hbox{and}  \hskip.5cm \displaystyle{{\bv'}_*  =  \bv_* + \bk (\bk \cdot \bq)},
\end{equation}
where $\bk$ is the unit vector in the direction of the apse-line bisecting $- \bv$ and $\bv'$, while the whole set of interactions  which is obtained by integration of $\bk$ over the domain:
$$
{\cal K} = \big\{\bk \in \mathbb{S}^{d-1} : \hskip.3cm ||\bk|| = 1, \hskip.3cm \bk \cdot \bq \hskip.1cm  \geq 0\big\}.
$$
Here $\mathbb S^{d-1}=\{\,\bk\in\mathbb R^d:\|\bk\|=1\,\}$ denotes the unit sphere in $\mathbb R^d$, i.e. the set of all unit vectors (directions).

We have chosen the term \textit{interaction} rather than \textit{collision}, which is often used in the literature, as particles do not really collide as repulsive forces modify their trajectories similarly to a collision, but not precisely as in the case of billiard balls.

\vskip.2cm \noindent \textbf{Remark 4.1} \textit{The dynamics of the  interactions leads to dynamical equations, i.e. the celebrated Boltzmann equation, which  admits a unique equilibrium configuration, i.e. the \textit{Maxwellian equilibrium distribution}, in which the speed is defined in the whole space $\RR_+$, while the distribution asymptotically tends to zero, when the speed tends to infinity:
\begin{equation}\label{Maxwellian}
f_e (\bx, \bv) = \frac{\rho(\bx)}{{\left[2 \pi (k/m) \Theta(\bx)\right]}^{3/2}}
\exp \bigg\{- \frac{|\bv - \bU(\bx)|^2}{2 (k/m) \Theta(\bx)} \bigg\},
\end{equation}
where $k$ is the Boltzmann constant, and $\Theta$ is the temperature given, for a mono-atomic gas, by the approximation of a perfect gas, namely:
${\cal E} = \frac{3}{2} \, k\, \Theta$, where ${\cal E}$ is the kinetic energy. Furthermore $\rho$ is the local density and $\bU$ is the mass velocity.}

\vskip.2cm \noindent \textbf{Remark 4.2} \textit{The tendency to equilibrium is ensured by the \textit{kinetic entropy}, namely the $H$-Boltzmann functional:
\begin{equation}\label{H}
H[f](t) = \int_{\RR^3 \times \RR^3} f\,\log (f)(t, \bx, \bv)\, d\bx\,d\bv,
\end{equation}
which, in the spatially homogeneous case, is monotonically decreasing along the solutions and is equal to zero at $f=f_e$. Therefore the Maxwellian distribution with parameters $\rho$, $\bU$, and
$\Theta$ minimizes $H$ (this is the celebrated $H$-Theorem~\cite{[CIP93]}).The absolute value of $H[f_e]$ depends on the normalization of $f$; some conventions shift $H$ by an additive constant so that $H[f_e]=0$. What matters for the $H$-Theorem is that the Maxwellian uniquely minimizes $H$.}

\vskip.2cm
The solution of the Boltzmann equation provides the macroscopic quantities by low order models. Suppose that $|\bv|^r \, f(t, \bx, \bv) \in L_1(\RR^3),  \hskip.2cm   \hbox{for} \quad r = 0, 1, 2, \ldots$, then macro-scale quantities are obtained by weighted moments as follows:
\begin{equation}\label{moments}
M_r = M_r[f](t, \bx)= \int_{\RR^3} m\, \bv^r \,  f(t, \bx, \bv)\,d\bv,
\end{equation}
where $m$  is the mass of the particles, $r=0$ corresponds  to the local density $\rho(t, \bx)$, $r=1$ to the linear momentum $\bQ(t, \bx)$,
while $r=2$ to the mechanical energy. Therefore, the local density and mean velocity are computed as follows:
\begin{equation}\label{density}
\rho =\rho(t,\bx) = m \int_{\RR^3} f(t, \bx, \bv)\,d\bv,
\end{equation}
\begin{equation}\label{mean}
\bU = \bU(t, \bx)= \frac{1}{\rho(t, \bx)} \int_{\RR^3} m\, \bv \,  f(t, \bx, \bv)\,d\bv,
\end{equation}
while the kinetic translational energy is given  by the second order moment
\begin{equation}\label{energy}
{\cal E} = {\cal E}(t, \bx)= \frac{1}{2 \rho(t, \bx)} \int_{\RR^3} m\,|\bv|^2 \,  f(t, \bx, \bv)\,d\bv,
\end{equation}
where $|\bv|$ is the velocity modulus or speed. The energy expression (\ref{energy}) can be related,  equilibrium conditions, to the local temperature based on the principle of repartition of energy, which is however valid only in the said conditions.
The energy expression (4.8) can be related to the local temperature under equilibrium conditions, based on the principle of energy repartition, which, however, is valid only under those conditions.

Some authors have used the so-called discrete Boltzmann equation as a reference for kinetic theory, see~\cite{[Gatignol],[MP91]}. Indeed, the framework provided by the discrete velocity model is useful for modelling the dynamics of active particles, as shown in some applications such as models of vehicular traffic, see~\cite{[CDF07]}, and crowd dynamics, see~\cite{[KQ19]} and~\cite{[KOOQ21]}.

\subsection{Interactions of active particles}\label{Sec:4.2}

Following the presentation in the previous subsection, we consider the dynamics of active particles in the KTAP approach. We refer to \cite{[BBD21]} and to the review paper~\cite{[BBD21]} for the theory, which proposes the following rationale:
\begin{itemize}
\vskip.2cm  \item  Interactions  modify the microscopic state, i.e. activity, velocity direction, and speed, which depends on the  microstate and distribution function of the a-particles  in the interaction domain.

\vskip.2cm \item The a-particles interacting in the system can be distinguished in the following statistically interpreted particles: \textit{Test particles}, which are representative of the whole system. Their microscopic state is $\bw = \{\bx, \bv, \bu\}$. \textit{Field particles}, which interact with each other.  Their microscopic state is $\bw_* = \{\bx_*, \bv_*, \bu_*\}$. \textit{Candidate particles}, which are the specific field particles that are likely to acquire the microscopic state of the test particle after interacting with the field particles. Their microscopic state is $\bw^* = \{\bx^*, \bv^*, \bu^*\}$. The distribution function of the test AP, which defines the state of the system, is $f(t,\bw)$. The distribution functions of the field and candidate a-particles are $f_* = f(t,\bw_*)$ and $f^* = f(t,\bw^*)$, respectively.

\vskip.2cm \item Interactions of \textit{test} and \textit{candidate} with \textit{field} particles take place within  \textit{the local interaction domain}, which is defined by a conical region $\Omega$, whose vertex is in $\bx$, the radius is defined by $R$, while the angle $\theta$ defines the width of the cone. In this domain, test or candidate a-particles learn  the individual and collective state of the field particles.

\vskip.2cm \item Interactions occur with a \textit{interaction rate}, which models the interactions of test or candidate particles with field particles. The interactions are  not reversible and not local. In fact, active particles are sensitive to other particles at a distance.

\vskip.2cm \item The states of  \textit{test} and \textit{candidate} are modified according to the \textit{transition probability density}, which  models the probability that a candidate particle at $\bx$ with state $\{\bv_*, \bu_*\}$ shifts  to the state of the test particle $\{\bv, \bu\}$ due to the interaction with field particles with state  $\bw^*$ in $\Omega$.

\vskip.2cm \item The system is not numerically conservative. In fact, encounters can generate proliferative and destructive dynamics, which are described by the \textit{proliferative and destructive probability}.
\end{itemize}

\vskip.2cm \noindent  \textbf{Remark 4.3.} \textit{Interactions can depend not only on the state of the interacting pairs, but also on the distribution functions. We define \textbf{linear interactions} as those in which the rate and output of the interactions depend only on the microscopic state of the interacting particles, while we define \textbf{nonlinear interactions}  those in which the state of the system also depends on the state of the distribution function. The interaction domain depends on $f$, i.e. $\Omega = \Omega [f]$.}
\vskip.2cm

The following notations are used for nonlinear interactions:
\begin{equation}\label{formal-A}
\eta[f](\bw, \bw_*), \hskip.5cm \Omega [f](\bw, \bw_*), \hskip.5cm \cP[f](\bw, \bw_*), \hskip.5cm \cD[f](\bw, \bw_*),
\end{equation}
for encounters between test and field particles, assuming that proliferation and destruction occur in the state of the test particles, and
\begin{equation}\label{formal-B}
\eta[f](\bw*, \bw_*), \hskip.5cm \Omega [f](\bw^*, \bw_*), \hskip.5cm \cA[f](\bw_* \to \bw; \bw^*, \bw_*),
\end{equation}
for encounters between candidate and field particles that cause the candidate to transition to the test particle state. These quantities correspond to the interaction rate, transition probability density, sensitivity range, proliferation probability, and destruction probability, respectively. The notation $[f]$ denotes the dependence on the distribution function of the interacting a-particles.

The above description has been developed for a single FS. If the overall system comprises a number $n$ of FSs, that can be denoted by the subscript $i = 1, \ldots, n$, the terms in Eqs.~(\ref{formal-A}) and~(\ref{formal-B}) can be denoted by the subscript $ij$, which indicates interactions between the i-th and j-th FSs, with $j = 1, \ldots, n$. The following notations can be used to denote the interactions involving different functional subsystems:
\begin{equation}\label{formal-C}
\eta_{ij}[f_i, f_j](\bw, \bw_*), \hskip.1cm \Omega [f_i, f_j](\bw, \bw_*),  \hskip.1cm  \cP_{ij}[f_i, f_j](\bw, \bw_*), \hskip.1cm \cD_{ij}[f_i, f_j](\bw, \bw_*),
\end{equation}
and
\begin{equation}\label{formal-D}
\eta_{ij}[f_i, f_j](\bw*, \bw_*), \hskip.5cm \Omega [f_i, f_j](\bw^*, \bw_*), \hskip.5cm \cA_{ij}[f_i, f_j](\bw_* \to \bw; \bw_*, \bw^*).
\end{equation}

The notation $\Omega [f_i f_j](\bw, \bw_*)$ needs some additional considerations to make precise the way by which this term is related to
the  conjecture in~\cite{[BCC08]}, according to which each a-particle feels only a fixed number of particles, in the continuous case we consider a critical density $\rho_c$.

Additional considerations  make precise the nonlinearity features in $\Omega$. In fact, accepting the conjecture in~\cite{[BCC08]}, if $\rho_c$ is the critical density necessary to take a decision for the dynamics, some considerations can be done in the case of plane motion.
Let us consider an arc of circle with axis along the velocity vector, let $\theta^+ = \theta^-  = \theta$  be the two angles that define the semi-amplitudes that define the domain, and $R_s$ is the radius that defines $\Omega[f](\bx)$. Then, the following relation defines $R_s$:
\begin{equation}\label{domain}
\rho_c = \int_{\Omega[f]} f(t, \bx, \bv) \, d\bx, \hskip.2cm \Rightarrow \hskip.2cm R_s[f], \hskip.2cm \Rightarrow \Omega = \Omega[f],
\end{equation}
that defines the nonlinear dependence of $\Omega$ with respect to $f$. As an example Eq. \ref{domain} is nonlinear and can be solved only if the geometry of $\Omega$ is defined. In this case the unknown is $\mathbb R$.
If polar coordinates can be used, $\bx = \{r \, \cos \theta, r\, \sin \theta \}$, so that $d \bx = r\, dr \, d\theta$. Analogous calculations can be done for different angles of the amplitudes $\theta^+$ and  $\theta^-$ or considering a conical domain in space. On the other hand, calculations for different interacting subsystems have not been developed yet.

\vskip.2cm \noindent  \textbf{Remark 4.4.} \textit{Real-world applications for self-propelled particles suggest to also consider the visibility domain, which is denoted by the same geometry, but with radius $R_v$. If $R_v \geq R_s$, $\Omega$ is calculated with $R_s$. On the other hand, if $R_v < R_s$, then $\Omega$ is computed with $R_v$. In general, both sensitivity and visibility can induce asymmetric sensitivity angles, i.e., $\theta^+$ and $\theta^-$. In this case, an open challenging problem is to understand how the dynamics of ``learning'' $\to$ ``decision making'' are modified.}
\vskip.2cm

We refer to the open access paper~\cite{[BBD21]}, see Section 3, to report, for sake of completeness, the structures, which generate model of real-world systems.

These models are constructed by incorporating interaction models obtained through a phenomenological and multiscale interpretation of the physical system under study.
by inserting into them, the interaction models, which can be obtained by a phenomenological and multiscale interpretation of the physical system object of the modelling approach.
\begin{equation}\label{MultiFS}
\partial_t f_i(t, \bx, \bv, \bu) + \bv \cdot \nabla_\bx \, f_i(t, \bx, \bv, \bu) = (\cG_i[\bbf] - \cL_i[\bbf])(t, \bx, \bv, \bu),
\end{equation}
where the terms $\cG_i$ and $\cL_i$ represent the gain and loss, respectively, of the number of a-particles entering and leaving the elementary volume $d\bx d\bv d\bu$ of the microscopic state space at time $t$.

Their expression is as follows:
\begin{eqnarray} \label{G-MultiFS}
&&\cG_i[\bbf] = \sum_{j=1}^n \int_{\Gamma \times {D_\bv} \times {D_\bu}} \eta_{ij}[f_i, f_j](\bx, \bx^*, \bv_*,\bv^*, \bu_*, \bu^*)\nonumber\\[2mm]
 &&  \hskip1cm \times \cA_{ij}[f_i, f_j](\bv_* \to \bv, \bu_* \to \bu|\bx, \bx^*, \bv_*,  \bv^*, \bu_*, \bu^*) \nonumber \\[2mm]
&& \hskip2cm  \times  \, f_i(t, \bx, \bv_*, \bu_*) f_j(t, \bx^*,  \bv^*, \bu^*)\, d\bx^*\,d\bv_*\,d\bv^* \, d\bu_*\, d\bu^*,
\end{eqnarray}

and
\begin{eqnarray} \label{L-MultiFS}
&&  \cL_i[\bbf] = f_i(t, \bx, \bv,\bu) \sum_{j=1}^n \int_{\Gamma} \eta_{ij}[f_i, f_j](\bx,\bx^*, \bv,\bv^*,\bu, \bu^*)\nonumber \\[2mm]
 && \hskip2.5truecm \times \,f_j(t, \bx, \bx^*, \bv^*, \bu^*)\,d\bx^* \, d\bv^* \, d\bu^*,
\end{eqnarray}
where $\Gamma = \Omega \times {D_\bv} \times {D_\bu}$.

\vskip.2cm \noindent  \textbf{Remark 4.5.} \textit{Boundary conditions should provide a reflection model at the wall as in the case of kinetic models of classical particles. However, each a-particle feel the presence of the walls at distance and modify their trajectories accordingly. This topic is treated in~\cite{[BGQR22]}, see Sections 4 and 5.}

\vskip.2cm \noindent  \textbf{Remark 4.6.} \textit{Consider first the applications of behavioral kinetic theory to the modelling of self-propelled particles, which have focused mainly on the dynamics of crowds. We will focus on the recent literature, see~\cite{[BGQR22]}, which shows how the same modelling principles can be applied at all scales, microscopic, mesoscopic, and macroscopic, according to the key considerations of Hilbert's Sixth Problem, see~\cite{[Hilbert]}. In particular, it is shown how the modelling of interactions leads to the description of collective motion. These results have been reviewed in~\cite{[BLQRS23]}, based on various contributions for crowds, see~\cite{[AAK25],[ABKT23],[BHLY24],[KQ19],[KOOQ21],[KLMY25],[LZ22]}, and for the dynamics of swarms, see~\cite{[BH17]}}.

\subsection{Interactions in spatial homogeneity}\label{Sec:4.3}

All of the above considerations can also be applied to systems where space dynamics and velocity do not play a role. In this case, the distribution function depends only on time and activity. Thus, the distribution functions of the interacting particles are:
$$f(t, \bu), \hskip1cm f_*(t, \bu_*),  \hskip1cm  f^*(t, \bu^*).
$$

The interactions depend on the activity and the distribution functions. The expression of the interaction terms in Eq.(\ref{formal-C}, \ref{formal-D}) are obtained by simply replacing $\bw, \bw_*, \bw^*$ by $\bu, \bu_*, \bu^*$. For each component of the activity a domain $D_u = [0, 1]$ can be considered, where $u=0$ and $u=1$ correspond to the minimum and maximum real values of the activity.

\vskip.2cm \noindent  \textbf{Remark 4.5.} \textit{The above notations illustrate another difference between classical and active particle dynamics. In fact, the spatially homogeneous dynamics in kinetic theory corresponds to the dynamics of the velocity distribution, which tends homogeneously in space to the Maxwellian equilibrium. Then the dependent variable is $f = f(t,v)$ and the differential structure can be written as
$$
\partial_t f(t,v) = Q(f,f)(t,v).
$$
The dynamics is different for active particles because interactions can be developed by signals, so position and velocity do not have the same meaning as in the dynamics of classical particles. The study is restricted to a distribution function independent of space and velocity, while  the term \textit{vanishing variables} has been occasionally used.}
\vskip.2cm

The mathematical structure is here reported for a scalar variable, i.e.  for a $f = f(t,u)$. We consider a model which includes proliferative and destructive dynamics, see~\cite{[BBD21]}.
\begin{equation}\label{hspace-a}
\partial_t f_i(t, u) = \cG_i[\bbf](t,u) -  \cL_i[\bbf](t,u),
\end{equation}
where
\begin{eqnarray}\label{hspace-b}
&& \cG_i[\bbf](t,u) = \sum_{j=1}^n \,\int_{\Omega} \eta_{ij}(u,u^*)[f_i,f_j] \mathcal{A}_{ij}^i[f_i,f_j]\left(u_* \to u|u,u^*\right) f_i(t,u)f_j(t, u^*)\,du^* \nonumber\\[3mm]
&& \hskip.5cm + \sum_{j=1}^n \,\int_{\Omega} \eta_{ij}(u,u^*)[f_i,f_j] \mathcal{P}_{ij}^i[f_i,f_j]\left(u_* \to u|u,u^*\right) f_i(t,\bu)f_i(t,\bu^*)\, d\bu^*,
\end{eqnarray}
and
\begin{eqnarray}\label{hspace-c}
&&  \cL_i[\bbf](t,u) = \sum_{j=1}^n \, \int_{\Omega}  \eta_{ij}[f_i,f_k](u,u^*)\,f_i(t, u) f_j(t,u^*)\,du^*\nonumber\\[2mm]
&& \hskip1cm   +  \sum_{j=1}^n \,\int_{\Omega} \eta_{ij}[f_i,f_j](u, u^*)\, \mathcal{D}_{ij}[f_i,f_j]\, f_i(t, u)\,f_j(t, u^*)\, du^*.
\end{eqnarray}

Additional considerations may focus on the interaction domain in the case of scalar activity. For example, we can set $\Omega_(ij) = [0, 1] or D_u = [0, 1]$ if it is independent of the interaction populations. The dimensionless variable u is obtained by referencing the activity to the maximum observed value. An empirical definition of \textit{symmetric interactions} is when the sensitivity domain is $D_u^s = [0, 1]$, while asymmetric interactions occur when $D_u^s \subset [0, 1]$. The conjecture in \cite{[BCC08]} has not been studied mathematically in this particular case. Arguably, as a consequence, most applications use $D_u^s = D_u = [0, 1].   D_u^s = D_u = [-1, 1]$,
the activity can take negative values.

The first applications of the differential structures reported in Eqs.(\ref{hspace-a}-\ref{hspace-c}) have been focused on the mathematical description of the immune competition between tumor and immune cells, see~\cite{[BD06]}, in which  immune cells develop their defense abilities by progressing  from the state of innate immunity to that of active immunity.  In contrast tumor cells progress their activation up to metastatic competence and proliferate by invading the surrounding tissue. This seminal paper has inspired various authors as demonstrated  in the book~\cite{[BD06]}.
The biology of cancer phenomena  and  a general theory of immune competition are reported in the books~\cite{[WEI07]} and~\cite{[MF19]}, respectively.

A natural extension of the  above studies has been the search for differential models deemed to depict the dynamics of the SARS-CoV-2 pandemic. In particular, The study of the in-host dynamics, which has been studied through active particle methods, has provided a detailed description of the pathology including the modelling of the level of pathology and, consequently,  the level of infectivity in a heterogeneous population.  See references~\cite{[BBC20],[BBO22],[BK24]}, and~\cite{[B3EPT]} for a review of the literature and the further developments of the kinetic theory of active particles.

An important area of research where active particle methods have been applied is the modelling of social dynamics. A previous review paper critically analyzes the literature from the past decade; see~\cite{[DLO17]}. Recently, important progress has been made. This matter will be discussed in the next section, which focuses primarily on this topic.

\section{Towards research perspectives}\label{Sec:5}

Previous sections provided a detailed study of different types of interactions involving a-particles. This section demonstrates how the interactions' dynamics can be interpreted using the kinetic theory of active particles. This theory translates the dynamics into the distribution function of the a-particles' microscopic states.

We refer to the kinetic theory of active particles, which was recently reviewed in the paper~\cite{[BBL25]}, the latter of which also propose further developments. Mathematical models are obtained by inserting analytical interaction models into the differential structures reported in Section 4. These structures transfer dynamics from individual-based scales to the collective dynamics of large systems constituted by a number of interacting living beings.
Implementing differential models with initial and/or boundary conditions can create mathematical problems. Solving these problems yields the dynamics of the dependent variable, i.e., the aforementioned distribution function. Macroscopic observable quantities can be obtained through the weighted moments of this function.

More specifically, we are relating to the differential structures in equations (\ref{hspace-a}, \ref{hspace-b}) and (\ref{hspace-c}). Our main focus is on the study of behavioral economics in relation to social systems, see~\cite{[Thaler16],[TS09]}. This choice is also motivated by the fact that the study of problems involving space dynamics has already been studied quite exhaustively in~\cite{[BLQRS23]}.

Interest in new research frontiers is well-documented in the current literature. For example, the last chapter of~\cite{[BL12]} suggests moving beyond interactions limited to consensus dynamics, emphasizing the need to model heterogeneity in groups of interest. The topics presented in this book are critically interpreted in the essay~\cite{[ABG16]}. These problems have indeed been addressed in the literature, and most of these topics have received exhaustive replies, as documented in~\cite{[BD06]}. Furthermore, recent works on the interaction between mathematics and the natural sciences have identified new perspectives that deserve attention and can contribute to the quest for a mathematical theory.

Therefore, this section proposes key problems selected from various possible options. In particular:

\vskip.2cm \noindent -- The complex structure of interactions, with a main emphasis on nonlinearity.

\vskip.2cm \noindent -- The time evolution of interaction rules  and  multiple types of dynamics.

\vskip.2cm \noindent -- From learning dynamics  to interactive platforms in the natural sciences.
\vskip.2cm

The presentation is based on a variety of concepts and ideas.  This preliminary work establishes the foundation for developing a conceptual framework that explores the complex relationship between mathematics and economics. This framework is based on a study of the specific topics discussed in the following subsections. This critical study of the existing literature concludes with a forward look at challenging perspectives. Specifically, we offer philosophical considerations on the pursuit of a mathematical theory applicable to the topics discussed in this section.

\subsection{Multiple and nonlinear interactions}\label{5.1}

As stated in previous sections, the modelling of interactions should go beyond linear and binary interactions. Indeed, our study focuses primarily on living matter, for which this is a necessary requirement to capture the complexity of real, living-and hence, behavioral-systems. Each individual, or a-particle, modifies their state not only through binary encounters, but also by accounting for all a-particles, or field-particles, within the sensorial domain $\Omega$, defined in Section 4.

The definition of nonlinear interactions is given in Section 4, being specifically emphasized in Remark 4.3. This concept is somewhat related to topological interactions. Accordingly, we can state that nonlinearity means that the output of interactions depends not only on  the microscopic state of the interacting a-particle, but also on their state defined by the distribution function. Furthermore,  topological interactions imply that each a-particle interacts with all particles in its  sensitivity domain, i.e. $\Omega$. Within this domain interactions can be mean-field, but also non-symmetric. Conversely, a large part of the literature on kinetic theory methods, not only proposes models only based on binary interactions, but also models based on linear interaction rules.

In addition to these considerations, the last chapter of the book~\cite{[BL12]} proposes research perspectives and states that the literature generally focuses on consensus dynamics. This motivated the search for new ways to model interactions that go beyond binary and linear assumptions. The review paper~\cite{[ABG16]}, which addresses the attention of applied mathematicians, critically analyzes this statement and encourages them to explore a rich variety of interaction dynamics.

The perspectives in~\cite{[BL12]} and~\cite{[ABG16]} have been followed in various papers, as examples, an interesting analytical contribution has been given in~\cite{[LMT24]}, while different types of nonlinearities have been studied in~\cite{[DL14],[DL15],[DLM21]}.
Across the three papers, nonlinearities take different but complementary forms. In the wealth redistribution model, they arise from a dynamic threshold that endogenously switches interactions between cooperation and competition, creating a feedback loop between redistribution rules and the evolving wealth distribution. In the opinion dynamics model, nonlinearity is driven by consensus--dissent thresholds acting both at the level of individual opinion distances and at the level of node-to-node influence, with the network structure amplifying the push toward collective consensus. In the liquidity model for interbank networks, nonlinearities are introduced through stochastic cooperative--competitive games, where banks link formation depends on regulatory liquidity constraints and transition probabilities are further shaped by interest rates, generating complex systemic patterns from simple local incentives.

These concepts are illustrated in Figure 2, which shows the two possible output states that can result from an interaction between an a-particle and a particle in a different microscopic state.  One output increases the state of the alpha particle with the higher state and decreases the state of the alpha particle with the lower state, while the other output does the opposite.
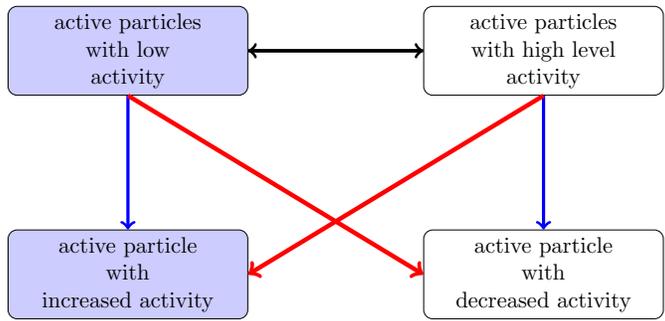
\begin{figure}[t!]
\begin{center}\scalebox{0.85}{
\begin{tikzpicture}[node distance = 4cm, auto]
 \node [block_long,  fill=white!] (High) {active particles  \\ with high level\\ activity};
 \node [block_long, left of=High,xshift=-2.5cm] (Low) {active particles  \\ with low \\ activity};
 \node [block_long, fill=white!, below of=High,yshift=.5cm] (Decreased) {active particle \\ with \\ decreased activity};
 \node [block_long, below of=Low, yshift=.5cm] (Increased) {active particle \\ with \\ increased activity};
 \draw [line width=.5mm, blue, ->] (High.south) -- (Decreased.north);
 \draw [line width=.5mm, blue, ->] (Low.south) -- (Increased.north);
 \draw [line width=.7mm, red, ->] (High.south) -- (Increased.east);
 \draw [line width=.7mm, red, ->] (Low.south) -- (Decreased.west);
  \draw[line width=.5mm, ->]  (High.west) -- (Low.east);
 \draw[line width=.5mm,->]  (Low.east) -- (High.west);
 \end{tikzpicture}}
\end{center}
\caption{\textbf{Blue color} direct arrows: Gain of the low level a-particles and loss of the high level a-particles.
\textbf{Red} color crossing arrows: Loss of the low level a-particles an d gain of the high level a-particles.}
\end{figure}

According to this draft, a-particles can choose one of two options based on their interpretation of the well-being target of interactions. The choice by which high-level a-particles contribute to low-level a-particles is sometimes defined as altruistic behavior. On the other hand, it may be imposed by governments or result from long-term thinking that concludes a society with strong inequalities is non-productive for open financial markets. Kahneman's book provides the conceptual framework for thinking fast and slow, which can be interpreted as the search for immediate or long-term well-being (see also \cite{KT79}, \cite{TK81}, \cite{KT92}).

This simple example corresponds to the above reasonings and some specific applications. In particular the dynamics of wealth distribution, see~\cite{[DL14]} or opinion formation, as an example see~\cite{[DL15]}.
The paper ~\cite{[DL14]} demonstrates that simple but nonlinear interaction rules among socio-economic agents can generate a wide range of emergent phenomena in wealth dynamics. A key modelling feature is a  threshold that may be externally fixed, representing government policy, or evolve endogenously through social competition. The resulting framework captures the wide spectrum of outcomes from almost unregulated individual exchanges to strongly policy-driven redistributive regimes, providing a natural explanation for the emergence of Pareto-like heavy tails in equilibrium distributions. The second one~\cite{[DL15]} investigates how network structures shape opinion dynamics by enhancing the emergence of consensus. Starting from simple nonlinear interaction rules of consensus and dissent, the model incorporates both local exchanges within each node and the influence of mean opinions across nodes. The analysis and simulations show that, while isolated populations may display persistent polarization, the presence of a network systematically favors convergence toward common opinions.

\subsection{On the dynamics of interaction's rules} \label{5.2}

Recent studies have developed a behavioral kinetic theory based on the assumption that individuals first learn from their neighbors and develop decision-making strategies based on a weighted sum of all available options. These decisions determine the dynamics. This theory was first proposed in ~\cite{[BBD21]} and was further developed in~\cite{[BBD21]} and~\cite{[BD06]}. Focusing specifically on economics, a schematic representation of the dynamics is as follows:
\begin{center}
\textbf{Economics} \, $\to$ \, \textbf{Learning} \,  $\to$ \textbf{Decision-Making}.
\end{center}

As mentioned above, our focus is on behavioral economics, which always interacts with other fields, such as politics and the social sciences. Then, we consider the model od brain-guided dynamics proposed in~\cite{[BD06]}. First, \textit{economics} selects the variables to be learned.  Then, \textit{individual a-particles} learn and, finally, make the decision.  This last act is made by maximizing an \textit{interpretation of the well-being} of individuals and/or groups of interest.These pioneering studies provide a conceptual framework that is worth further developing in the quest for a mathematical theory. The following points can contribute to this challenging goal.

\begin{itemize}

\vskip.2cm \item \textit{Learning dynamics} should address the specific features of each specific social-economic system. Then, the social sciences and economics are deemed to identify the variables at the microscopic scale (the state of the a-particles) and macroscopic scales (obtained by moments of the distribution function), which play an important role  in the subsequent decision-making. Thus, this step is science-aided.

\vskip.2cm \item The previous scientific study is also considered an elaboration of a theory supported by validated differential systems of possible choices for decision-making. This theory is brain-guided to maximize individual well-being related to the specific dynamics under consideration. This choice depends on how much each a-particle has learned.

\vskip.2cm \item Well-being corresponds to an individual's vision, which is heterogeneous and evolves over time in an artificial world with developing dynamics. According to Kahneman's theory of short and long thinking, this vision, i.e., the utility function, can be interpreted as such: see~\cite{[KA12]}. The time dependence of the utility function relates to the  Herbert A. Simon's theory of the artificial world, see~\cite{[Simon1965],[Simon1978],[Simon2019]}. For a mathematical-philosophical interpretation, refer to~\cite{[BE24]}.

\end{itemize}

\begin{figure}[t!]
\begin{center}\scalebox{0.85}{
\begin{tikzpicture}[node distance =3.0cm, auto]
\node [block_long] (Physics) {\textbf{A}: Study  of the\\ (Economics) \\ of a specific system};
\node [block_long, below of=Physics] (Learning) {\textbf{B}: Learning \\ interaction variables \\ microscopic \\ and macroscopic};
\node [block_long, left  of=Learning,xshift=-2.5cm] (Learn-micro) {\textbf{B-left}: Economics and \\ Brain guided \\ learning \\ microscopic variables};
\node [block_long, right of=Learning,xshift= 2.5cm] (Learn-macro) {\textbf{B-right}: Economics and \\ Brain guided \\  learning \\ macroscopic variables};
\node [block_long, below of=Learning] (Decision) {\textbf{C}: Brain guided \\ decision-making \\ toward differential\\  collective dynamics };
\draw [line width=.7mm, ->] (Physics.south) -- (Learning.north);
\draw [line width=.7mm, blue, ->] (Physics.west) -- (Learn-micro.north);
\draw [line width=.7mm, blue, ->] (Physics.east) -- (Learn-macro.north);
\draw [line width=.7mm, blue, ->] (Learning.south) -- (Decision.north);
\draw [line width=.7mm, blue, ->] ([yshift=0.3cm] Learn-micro.east) -- ([yshift=0.3cm] Learning.west);
\draw [line width=.7mm, blue, ->] ([yshift=-0.3cm] Learning.west) -- ([yshift=-0.3cm] Learn-micro.east);
\draw [line width=.7mm, blue, ->] ([yshift=0.3cm] Learn-macro.west) -- ([yshift=0.3cm] Learning.east);
\draw [line width=.7mm, blue, ->] ([yshift=-0.3cm] Learning.east) -- ([yshift=-0.3cm] Learn-macro.west);
\draw [line width=.7mm, blue, ->] (Learn-micro.south) -- (Decision.west);
\draw [line width=.7mm, blue, ->] (Learn-macro.south) -- (Decision.east);
\end{tikzpicture}}
\end{center}
\begin{center}
\caption{Brain-guided learning and decision-making}\label{fig1}
\end{center}
\end{figure}
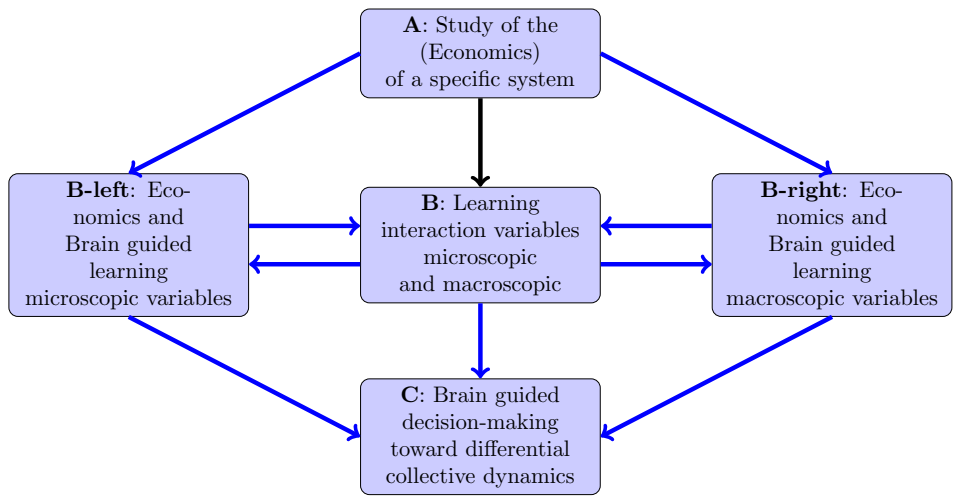

Additional considerations can refer to a specific features of social and economic systems, Indeed, different types of dynamics occur. Therefore, the following additional points should be considered.

\begin{itemize}

\vskip.2cm \item Current research in complex systems involves modelling various interactive dynamics in the fields of natural sciences and economics. For example, it includes  epidemics, social dynamics, and economics, see~\cite{[Aguiar],[B3EPT],[BDP20],[BK24]}.

\vskip.2cm \item In general, the different dynamics are  distributed heterogeneously within each functional subsystem, see~\cite{[APZ19]}. These can be a convex combination of the two dynamics as studied in the dynamics of wealth distribution, see~\cite{[DL14]} or opinion formation~\cite{[DL15]}. Different activities can characterize models of learning dynamics, see~\cite{[BDG16],[BDG16B],[BD19]}.

\vskip.2cm \item According to~\cite{[BDL24]} and ~\cite{[BD06]}, the draft of Figure 3 mentions that economics and the brain should guide learning and decision dynamics.  This perspective relates to scientific machine learning, which is somewhat inspired by Yann LeCun, see~\cite{[BLH21]} and~\cite{[CUN24]}.

\vskip.2cm \item Relating to modelling the cardiovascular system and  computational problems, see~\cite{[Alfio]}, while optimization algorithms for reinforcement are treated in various articles as documented in the bibliography of reference~\cite{[BHH25],[YLM22]}, see also~\cite{[Fornasier]}. The conceptual framework for achieving this objective is the physical interpretation of complexity, see~\cite{[Parisi23]}.

\vskip.2cm \item The study of complexity can focus specifically on economics, as discussed in references~\cite{[Arthur],[DR2019]} and~\cite{[Dosi2023]}, the latter of which relates to a broad framework of applications. A study of the analogy and conceptual differences between physics and economics are studied in~\cite{[JL21]}.

\end{itemize}

\subsection{On a quest to a mathematical theory}\label{5.3}

This paper naturally concludes with a quest for a mathematical theory of socio-economic systems. Specific questions relating to the interaction dynamics of well-defined applications were presented in previous sections. However, the topics proposed below do not directly lead to such a theory. Instead, they can solve interesting problems by interpreting specific systems phenomenologically, particularly with regard to modelling interactions. Therefore, they do not yet fit within the general framework of a theory.

Hence, this paper does not claim to have invented a theory. Instead, we propose and critically discuss three sample key topics that are necessary for the implementation of a theory. These topics have been selected based on the authors' research experience and forward-looking perspectives. Specifically, we consider the following topics:

\vskip.2cm \noindent $\bullet$ \textbf{Reasonings on the concept of mathematical theory:} As a mathematical theory, we consider a general differential system that is suitable for describing the temporal and spatial dynamics of the variables that are supposed to describe the state of social-economic systems.  Specifically, we refer to the system's dependent variables. This mathematical structure should provide the conceptual framework for deriving specific models. As demonstrated in previous sections, this involves deriving analytical models of interactions and incorporating them into the aforementioned structures.

Our approach considers systems in which social dynamics and economics interact. Examples of this type of multiple dynamics have been discussed in our paper, as examples, see~\cite{[DLO17]} and~\cite{[Dosi2023]}. Other contributions are reviewed in~\cite{[DLO17]} and~\cite{[CZR24]}.  A celebrated reference, see~\cite{[AR06]} focused on the interaction between politics and economics. Multiple interactions can involve social dynamics and epidemics, see~\cite{[AAF20]}. In the case of multiple interactions, the  activity variable must be a vector whose components have a reciprocal influence. Similarly, one can consider the different types of social dynamics as in the study of  language  interaction and evolution, (see for example, \cite{[BDMR20]}), as well as the different  economic systems competing within all nations.

\vskip.2cm  \noindent  $\bullet$  \textbf{From learning dynamics to decision-making:}  An important step is to develop the mathematical tools proposed in previous sections into a framework for scientific machine learning.  The suggestions in~\cite{[BDL24]} can contribute to this goal. The key idea is that the dynamics of individuals (the a-particles in our case) take place in two steps: first learning, then decision-making. This rationale has guided this essay, in which modelling interactions is a key part of the whole process.

Starting from~\cite{[BDG16]}, differential systems have been derived using kinetic theory methods to model collective learning. These systems have been applied to model different systems, as reviewed in~\cite{[BD06]}, where the first step towards scientific machine learning was outlined by linking learning and decision-making to a phenomenological model of the brain and physics-aided dynamics. The analogies between economics and physics, see reference~\cite{[JL21]} can contribute to model the aforementioned complex dynamics.

\vskip.2cm  \noindent  $\bullet$  \textbf{Validation of models within a multiscale framework.} Our paper presents a study that aims to understand
how dynamics at the microscopic scale are transferred by collective dynamics. This section mainly focuses on social dynamics and economics.  One challenging perspective is to develop a multiscale vision by refining the dynamics at a lower scale and deriving macroscopic models using asymptotic methods from the underlying microscopic description. This study was developed for biological systems, see references~\cite{[BC22]} and~\cite{[BKZ24]}, but  extending it to social dynamics remains an open problem. However, the conceptual framework is that of the 6th Hilbert's problem~\cite{[Hilbert]}.

Recent studies of biological systems, particularly those developed for modelling the competition between progressing viruses and the immune system (see, for example, references~\cite{[BK24]} and ~\cite{[BK25]}), suggest investigating the dynamics at a scale lower than that of cells. Indeed, sub-particles produced by cells are used to communicate. Some analogies can be drawn with the dynamics of social systems, in which individual entities (i.e. a-particles) communicate using visual and vocal systems, including the various media offered by technology.
\vskip.2cm

The three specific suggestions proposed above are not conclusive for this paper. Indeed, Yann LeCun proposed the challenging objective of deriving a differential system for brain-aided learning and decision-making for social and economic systems, see references~\cite{[BLH21]} and~\cite{[CUN24]}, where the topic is communicated  in a colloquial style. Further reasoning is available in reference~\cite{[WCL22]}. This topic definitely deserves further work, as it opens the door to scientific machine learning.

The study of crowd dynamics is an example of a research topic that has yielded preliminary results., see references~\cite{[BDL24]} and~\cite{[BD06]}. These articles encourage further research in this area of study. In particular, the first paper notes that studying this topic is important for addressing safety issues and that training-predictive platforms can facilitate dialogue between scientists and end users. The latter paper revisits mathematical modeling of learning dynamics to support decision-making and, more generally, the mathematical description of complex systems.

\section*{Acknowledgments}{The first author acknowledges the King’s Business School PhD programme at King’s College London.}

\input{Ref-MDLL-Rrevision}

\end{document}

%% file: Ref-MDLL-Rrevision.tex
%
%
%